\documentclass[a4paper,11pt]{article}
\usepackage{pos}
\usepackage{latexsym}
\usepackage{hyperref}
\usepackage{graphicx}
\usepackage{epstopdf}
\usepackage{bm}
\title{Isovector nucleon form factors from 2+1-flavor dynamical domain-wall lattice QCD at the physical mass}
\ShortTitle{DWF nucleon form factors}

\author*[a]{Shigemi Ohta}

\affiliation[a]{Institute of Particle and Nuclear Studies, High-Energy Accelerator Research Organization (KEK), Tsukuba, Ibaraki 305-0801, Japan}

\emailAdd{shigemi.ohta@kek.jp}

\abstract{
\vspace{-135mm}\parbox{\textwidth}{\flushright\large\rm \hfill KEK-TH-2457}\vspace{128mm}
Nucleon isovector form factors calculated on a 2+1-flavor domain-wall-fermions ensemble with strange and degenerate up and down quarks at physical mass and lattice cut off, $a^{-1}$, of about 1.730(4) GeV, are reported.  The ensemble was generated jointly by RBC and UKQCD collaborations with a spatial extent of $48a$ or about 5.5 fm.  The form factors are calculated in collaboration with LHP collaboration as well.  The resulting shape parameters of the form factors, such as vector-charge mean squared radius, $\langle r_1^2\rangle$, or anomalous magnetic moment, $F_2(0)$ appear less dependent on possible excited-state contaminations than the corresponding charges.  Preliminary estimates are $\langle r_1^2\rangle \sim 0.142(13)\, \mbox{fm}^2$ and $F_2(0) \sim 3.22(8)$.
}

\FullConference{%
The 39th International Symposium on Lattice Field Theory,\\
8th-13th August 2022,\\
Rheinische Friedrich-Wilhelms-Universität Bonn, Bonn, Germany
}

\begin{document}
\maketitle



We report the isovector vector- and axialvector-current form factors of nucleon calculated using a 2+1-flavor dynamical domain-wall fermions (DWF) numerical lattice quantum chromodynamics (QCD) ensemble with lattice cut off, \(a^{-1}\), of about 1.730(4) GeV and physical light- and strange-quark mass.
DWF lattice QCD preserves continuum-like chiral and flavor symmetries at relatively large lattice spacing, such as 0.1 fm.
RIKEN-BNL-Columbia (RBC) and UKQCD collaborations have been jointly generating dynamical 2+1-flavor DWF numerical lattice-QCD ensembles for over a decade now \cite{RBC:2006jmm,RBC:2007yjf,RBC:2007bso,RBC:2008cmd,RBC-UKQCD:2008mhs,RBC:2010qam,Aoki:2010pe,RBC:2012cbl,RBC:2014ntl,Boyle:2015exm}.
We have been working at physical mass for a while \cite{RBC:2014ntl,Boyle:2015exm}.
We have used some of these DWF ensembles for studying nucleon \cite{Yamazaki:2008py,Yamazaki:2009zq,Aoki:2010xg,Ohta:2011vv,Lin:2011vx,Lin:2012nv,Ohta:2013qda,Ohta:2014rfa,Syritsyn:2014xwa,Ohta:2015aos,Abramczyk:2016ziv,Ohta:2017gzg,Ohta:2018zfp,Ohta:2019tod,Abramczyk:2019fnf}.
We found a deficit \cite{Yamazaki:2008py} in calculated isovector axial charge, \(g_A\), in comparison with its experimental value \cite{Workman:2022ynf}.
About ten percent deficit of the calculated results with pion mass from about 420 MeV to 170 MeV had not moved much \cite{Ohta:2011vv,Lin:2011vx,Lin:2012nv,Ohta:2013qda,Ohta:2014rfa,Ohta:2015aos,Abramczyk:2016ziv,Ohta:2017gzg,Abramczyk:2019fnf} as we refined our analysis with lighter-mass ensembles.
Almost all other calculations confirmed this at similar lattice cuts off and quark mass \cite{Dragos:2016rtx,Bhattacharya:2016zcn,Liang:2016fgy,Ishikawa:2018rew,Chang:2018uxx}. 
Since then, more calculations at almost physical mass have been conducted, bringing the calculated values closer to the experiment \cite{Chang:2018uxx,Bhattacharya:2016zcn,Shintani:2018ozy,Hasan:2019noy,Harris:2019bih}, sometimes covering the experimental value within relatively large statistical and systematic errors.

However, our unitary DWF calculations with better chiral and flavor symmetries, and consequently smaller systematic and statistical errors, observe some deficits.
Statistical significance of these results ranges dependent on renormalizations used for the axialvector current\cite{Ohta:2018zfp,Ohta:2019tod,Ohta:2021ldu}:
From about three standard deviations with the renormalization obtained in the meson-sector calculation to about five standard deviations with the renormalization using the nucleon vector charge.
We note the corresponding vector charge calculation suggests possible contamination from nearby excited states \cite{Ohta:2018zfp,Ohta:2019tod,Ohta:2021ldu,Bar:2021crj,Bar:2021zds}, in contrast to earlier DWF calculations that did not find any evidence for such contamination \cite{Abramczyk:2019fnf,Ohta:2021ldu}.

On the other hand, the form factors calculated at finite momenta transfer statistically fluctuate more than the charges calculated at zero momentum transfer.
As a result, the possible contamination from excited states is less detectable.
In other words, such possible contaminations could be hidden by more significant statistical fluctuations in the shape parameters of the form factors such as mean squared radii or magnetic moments \cite{Ohta:2021ldu}.



The results presented here were calculated using the ``48I'' \(48^3\times 96\) 2+1-flavor dynamical M\"{o}bius DWF ensemble at physical mass with Iwasaki gauge action of gauge coupling, \(\beta=2.13\), or of lattice cut off of  \(a^{-1} = 1.730(4)\) GeV, jointly generated by the RBC and UKQCD collaborations \cite{RBC:2014ntl}.
In total, 130 configurations, separated by 20 MD trajectories in the range of trajectory number 620 to 980 and by 10 MD trajectories in the range of trajectory number from 990 to 2160, except the missing 1050, 1070, 1150, 1170, 1250, 1270, and 1470, are used.
Each configuration is deflated \cite{Clark:2017wom} with 2000 low Dirac eigenvalues.
The ``AMA'' statistics trick  \cite{Shintani:2014vja}, with \(4^4=256\) AMA sloppy samples unbiased by four precision ones from each configuration, is used.
Gauge-invariant Gaussian smearing  \cite{Alexandrou:1992ti,Berruto:2005hg} with similar parameters as in the past RBC nucleon structure calculations is applied to nucleon source and sink, separated in time by \(8 \le T \le 12\) lattice units.
We obtained a nucleon mass estimate of 947(6) MeV from this ensemble \cite{Ohta:2018zfp,Ohta:2019tod,Ohta:2021ldu}.


The nucleon isovector vector- and axialvector-current form factors are  experimentally measured in lepton elastic scatterings off, or  \(\beta\) decay of, or muon capture by nucleons:
\[\langle p| V^+_\mu(x) | n \rangle = \bar{u}_p \left[\gamma_\mu
F_1(q^2) - i \sigma_{\mu \lambda}q_{\lambda} \frac{F_2(q^2)}{2m_N} \right]
u_n e^{iq\cdot x},
\]
\[
\langle p| A^+_\mu(x) | n \rangle = \bar{u}_p
            \left[\gamma_\mu \gamma_5  F_{A}(q^2)
             +\gamma_5 q_\mu \frac{F_{P}(q^2)} {2m_N}\right]  u_n e^{iq\cdot x}.
\]
They are related to various important nucleon observables such as: mean-squared charge radii, \(\langle r_1^2\rangle\), through the expansion of the vector form factor, \(\displaystyle F_1(Q^2) = F_1(0) - \frac{1}{6} \langle r_1^2\rangle Q^2 + ...\), in terms of momentum transfer squared, \(Q^2 = |q^2|\),  or anomalous magnetic moment,  \(F_2(0)\), or isovector axial charge, \(g_A=F_A(0)=1.2754(13) g_V\)  \cite{Workman:2022ynf}, of nucleon that determines neutron life and nuclear \(\beta\) strengths that in turn determines nuclear abundance.
\begin{figure}[b]
\begin{center}
\includegraphics[width=.7\textwidth,clip]{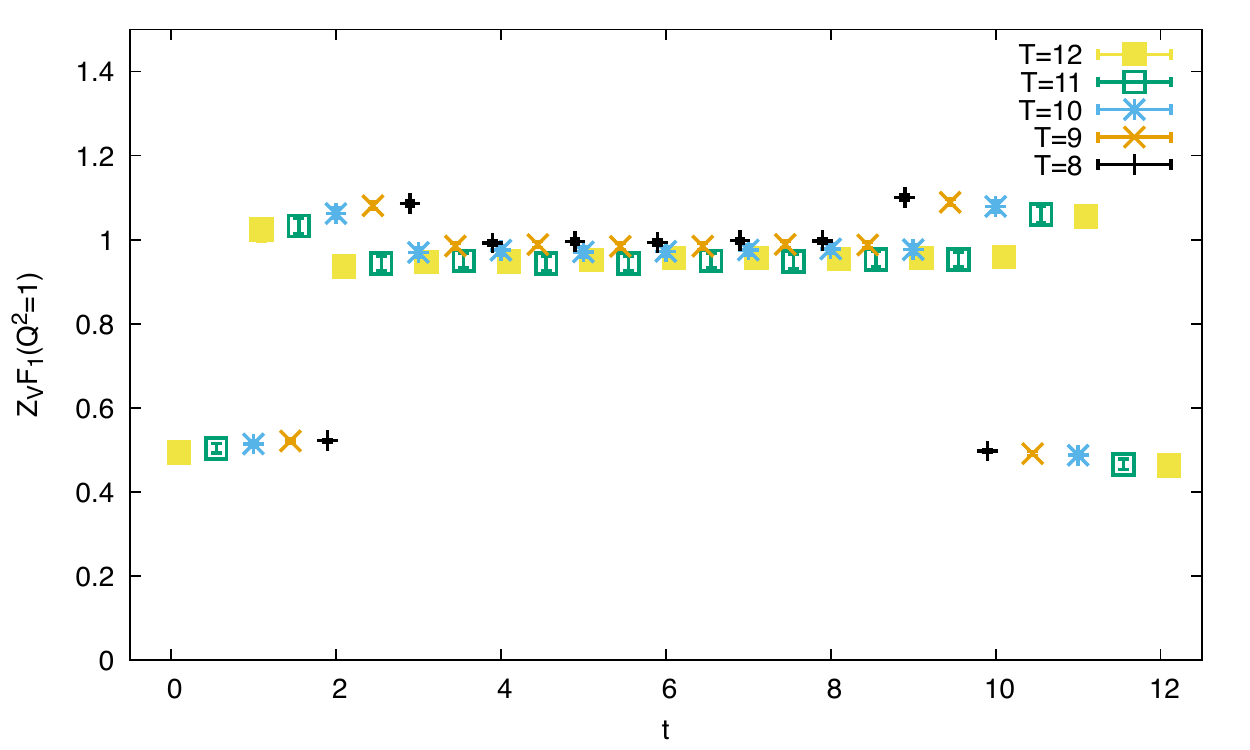}
\caption{
\label{fig:F1Q1}
Nucleon isovector vector form factor \(F_1\) with one lattice unit of momentum transfer squared, \(Q^2=1\), plotted against source-sink separations, \(T\), of 8, 9, 10, 11, and 12 lattice units.
The plateaux are well-defined and consistent with each other: suspected excited-state contamination is not detected.}
\end{center}
\end{figure}

To extract the form factors, we use the standard ratios,
\(\displaystyle 
\frac{C_{\rm 3pt}^{\Gamma,O}(t_{\rm src},t,t_{\rm snk})}{C_{\rm 2pt}(t_{\rm src},t_{\rm snk})}
\), of two-point,
\[
C^{(2)}(t_{\rm src},t_{\rm snk}) =
\sum_{\alpha, \beta}
\left(
\frac{1+\gamma_t}{2}
\right)_{\alpha\beta}
\langle
N_\beta(t_{\rm snk})\bar{N}_\alpha(t_{\rm src})
\rangle,
\]
and three-point,
\[
C^{(3)\Gamma, O}(t_{\rm src},t,t_{\rm snk}) =
\sum_{\alpha, \beta}
\Gamma_{\alpha\beta}
\langle
N_\beta(t_{\rm sink})O(t)\bar{N}_\alpha(0)
\rangle,
\]
correlators with a nucleon operator, \(N=\epsilon_{abc}(u_a^T C \gamma_5 d_b) u_c\). 
Plateaux of these ratios in time between the source and sink are obtained with appropriate spin (\(\Gamma=(1+\gamma_t)/2\) or \((1+\gamma_t)i\gamma_5\gamma_k/2\)) or momentum-transfer projections, which in turn give lattice bare value estimates for the expected values, \(\langle O\rangle\), for the relevant observables.
More specifically, for the form factors, ratios such as
\[
\frac{C^{(3)\Gamma, O}_{\rm GG}(t_{\rm src}, t, t_{\rm snk}, \vec{p}_{\rm src}, \vec{p}_{\rm snk})}
{C^{(2)}_{\rm GG}(t_{\rm src}, t_{\rm snk}, \vec{p}_{\rm src}, \vec{p}_{\rm snk})
} 
\times
\sqrt{
\frac{
C^{(2)}_{\rm LG}(t, t_{\rm snk}, \vec{p}_{\rm src}))
C^{(2)}_{\rm GG}(t_{\rm src}, t, \vec{p}_{\rm snk}))
C^{(2)}_{\rm LG}(t_{\rm src}, t_{\rm snk}, \vec{p}_{\rm snk}))
}
{
C^{(2)}_{\rm LG}(t, t_{\rm snk}, \vec{p}_{\rm snk}))
C^{(2)}_{\rm GG}(t_{\rm src}, t, \vec{p}_{\rm src}))
C^{(2)}_{\rm LG}(t_{\rm src}, t_{\rm snk}, \vec{p}_{\rm src}))}
}
\]
with point (L) or Gaussian (G) smearings, give plateaux dependent only on momentum transfer.
Further details can be found in our earlier publications, such as Ref.\ \cite{Yamazaki:2009zq}.

We use the source-sink separation, \(T\), from 8 to 12, following the earlier studies for isovector charges and couplings \cite{Ohta:2018zfp,Ohta:2019tod,Ohta:2021ldu}.
All the 147 three-momentum transfers \(\vec{Q}\)with \(Q^2 \le 10\) are included.
Note there is no such three-momentum with \(Q^2=7\) lattice units.
The results for the vector form factor at the minimum finite momentum transfer of \(Q^2=1\) in the lattice unit presented in Fig.\ \ref{fig:F1Q1} are encouraging:
Since the numbers from the shortest source-sink separation of \(T=8\), with the minor statistical fluctuations, are in agreement with the numbers from longer separations of \(T=9\), 10, 11, and 12 with successively more significant statistical fluctuations, we should be able to extract shape parameters such as mean square radii or magnetic moment from these form factors without detectable contamination from excited states.
As even the ground-state signals deteriorate beyond \(T\ge 11\), it is best to use the calculations with \(T \le 10\).

\begin{figure}[b]
\begin{center}
\includegraphics[width=.7\textwidth,clip]{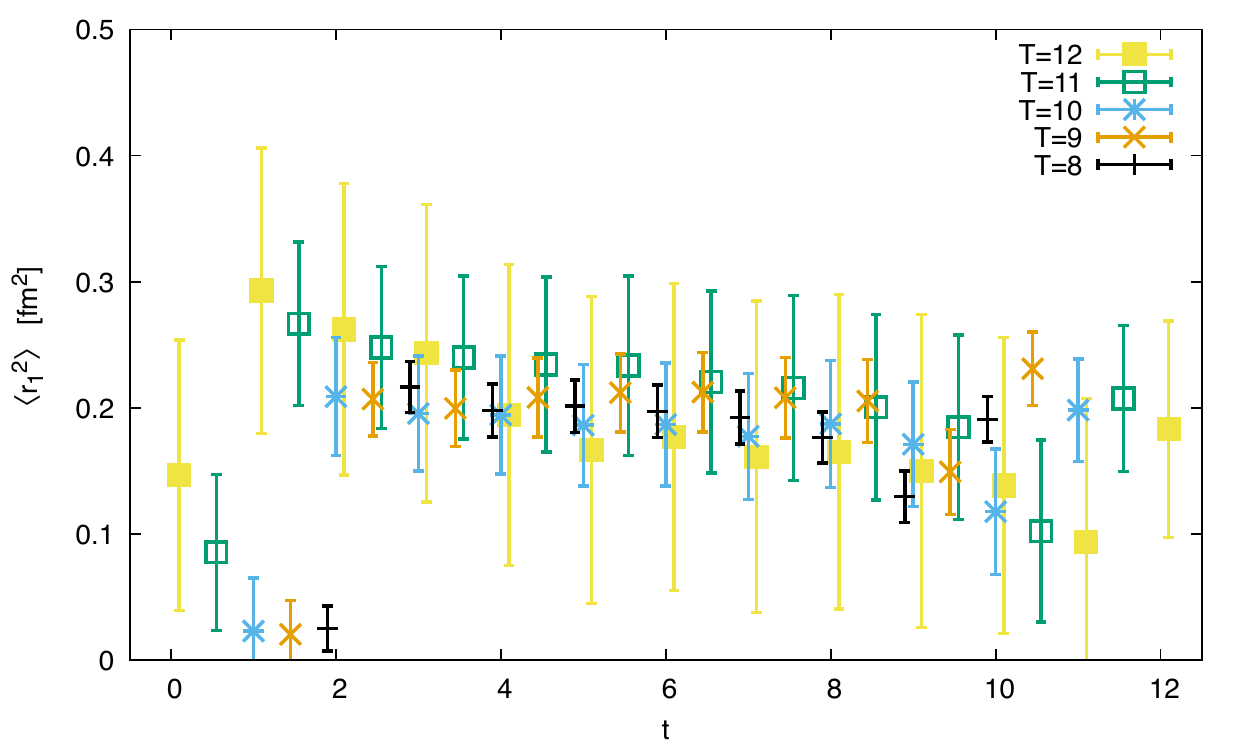}
\caption{
\label{fig:R1Q1}
The mean-squared radius, \(
\langle r_1^2\rangle = \frac{6 [F_1(Q^2=0) - F_1(Q^2=1)] }{Q^2=1}
\) does not seem to depend on the source-sink separation.
These average \(\sim 0.20(2) \mbox{fm}^2\) as compared to
experiment: \([(0.8409(4))^2+0.1161(22) = 0.8682(29)](\mbox{fm})^2\)}
\end{center}
\end{figure}

Indeed in the mean-squared charge radius, as defined by \(\displaystyle
\langle r_1^2\rangle = \frac{6 [F_1(Q^2=0) - F_1(Q^2=1)] }{Q^2=1}
\), in Fig.\ \ref{fig:R1Q1},
no excited-state contamination is detectable above the statistical errors:
As can be expected from the form factor values, \(F_1(Q^2=1)\) shown in Fig.\ \ref{fig:F1Q1}, the estimates from shorter separations such as \(T=8\) and 9 with less statistical fluctuations are in agreement with those from longer separations such as \(T=10\), 11, and 12 with larger fluctuations.

The whole shape of the vector form factor, \(F_1\), is presented in Fig.\ \ref{fig:F1scaled}:
\begin{figure}[tb]
\begin{center}
\includegraphics[width=.7\textwidth,clip]{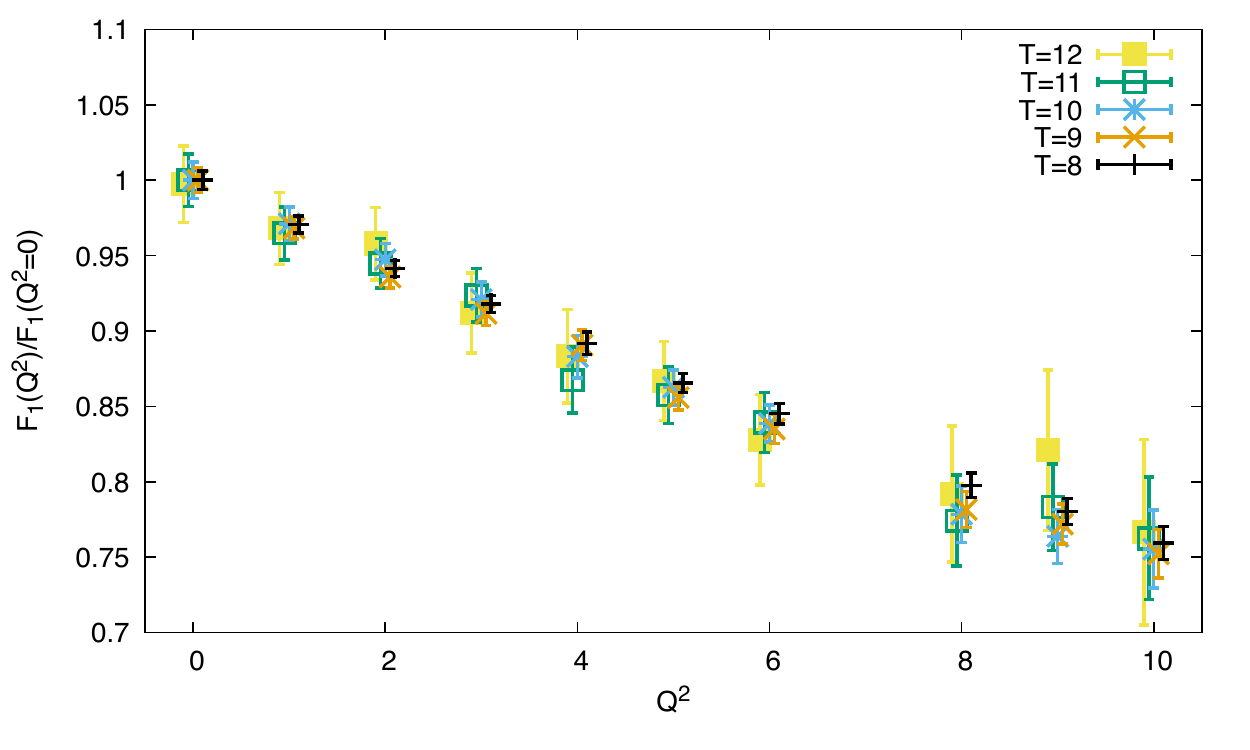}
\caption{
\label{fig:F1scaled}
Isovector vector form factor, \(F_1(Q^2)\), scaled by the corresponding charge, \(F_1(0)\).
The shape does not depend on source-sink separation \(T\).
Note that the values for \(Q^2=7\) are absent as there is no lattice momentum transfer combination with \(Q^2=7\).}
\end{center}
\end{figure}
The shape does not depend on source-sink separation \(T\) in the sense that the values calculated with shorter separations are well contained within the error bars of the values calculated with longer separations.
The suspected excited-state contamination is not detected in the whole shape either.

\begin{figure}[b]
\begin{center}
\includegraphics[width=.7\textwidth,clip]{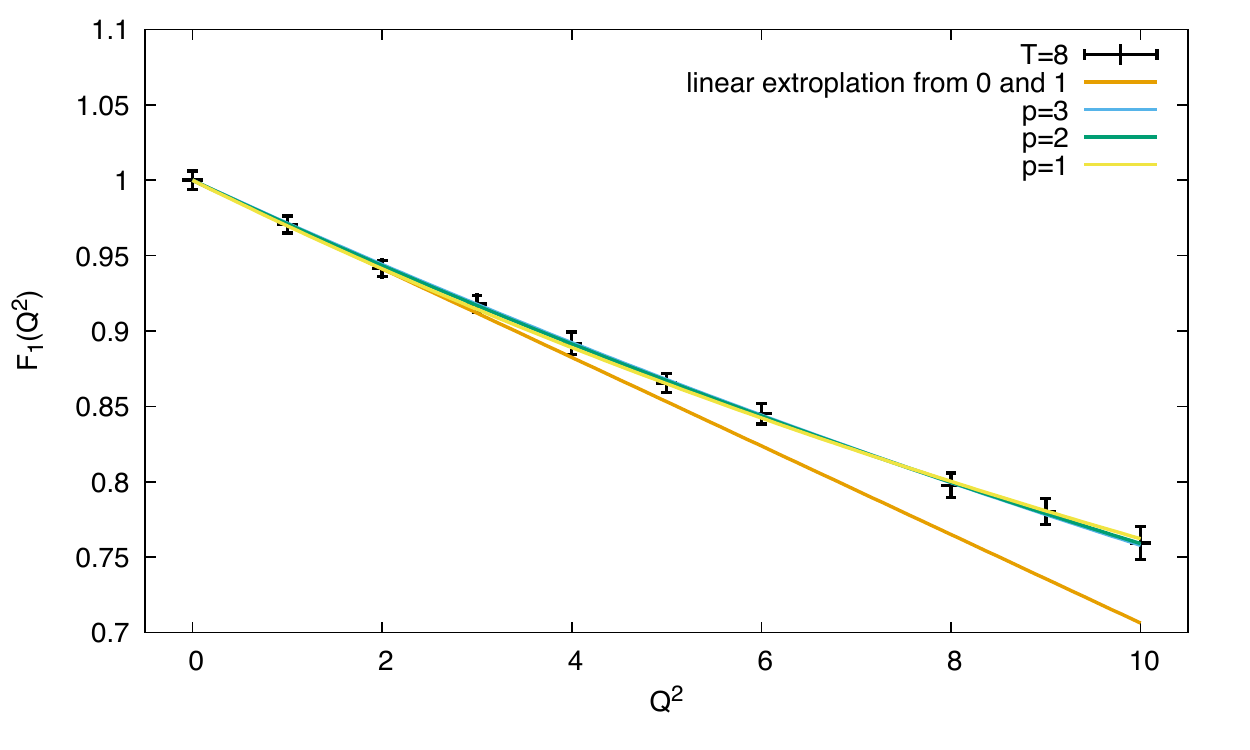}
\caption{
\label{fig:p123}
The isovector vector form factor is easily fit by a wide range of multipole forms, \(
F(Q^2) \sim F(0) \left(1+\frac{Q^2}{M_p^2}\right)^{-p}\).
Here vector form factor is compared with multipoles  \(p=1\),  2, and  3, as well as a linear fit using \(Q^2=0\) and 1 only.
All the fits result in similar estimates for the mean-square radius.
}
\end{center}
\end{figure}

This form factor is easily fit by a wide range of multipole forms, \((\displaystyle F(Q^2) \sim F(0) \left(1+\frac{Q^2}{M_p^2}\right)^{-p}\), not only with \(p=1\),  2, and  3  presented in Fig.\ \ref{fig:p123}
but also with \(p=4\), 5, 6, and 7, all resulting in \(\chi^2\) per degree of freedom below unity.
The resulting mean-squared charge radius estimates of \(\langle r_1^2\rangle = 6p/M_p^2 \sim 0.14\ \mbox{\rm fm}^2\) do not depend much on the multipolarity, \(p\), in a wide range of \(1 \le p \le 7\).
Nor do they differ from the linear extrapolation using only \(Q^2=0\) and 1.
However, the multipole form with  \(p\le 1/2\) or \(8 \le p\) does not work.

The induced tensor form factor, \(F_2\), is presented in Fig.\ \ref{fig:F2scaled}:
\begin{figure}[tb]
\begin{center}
\includegraphics[width=.7\textwidth,clip]{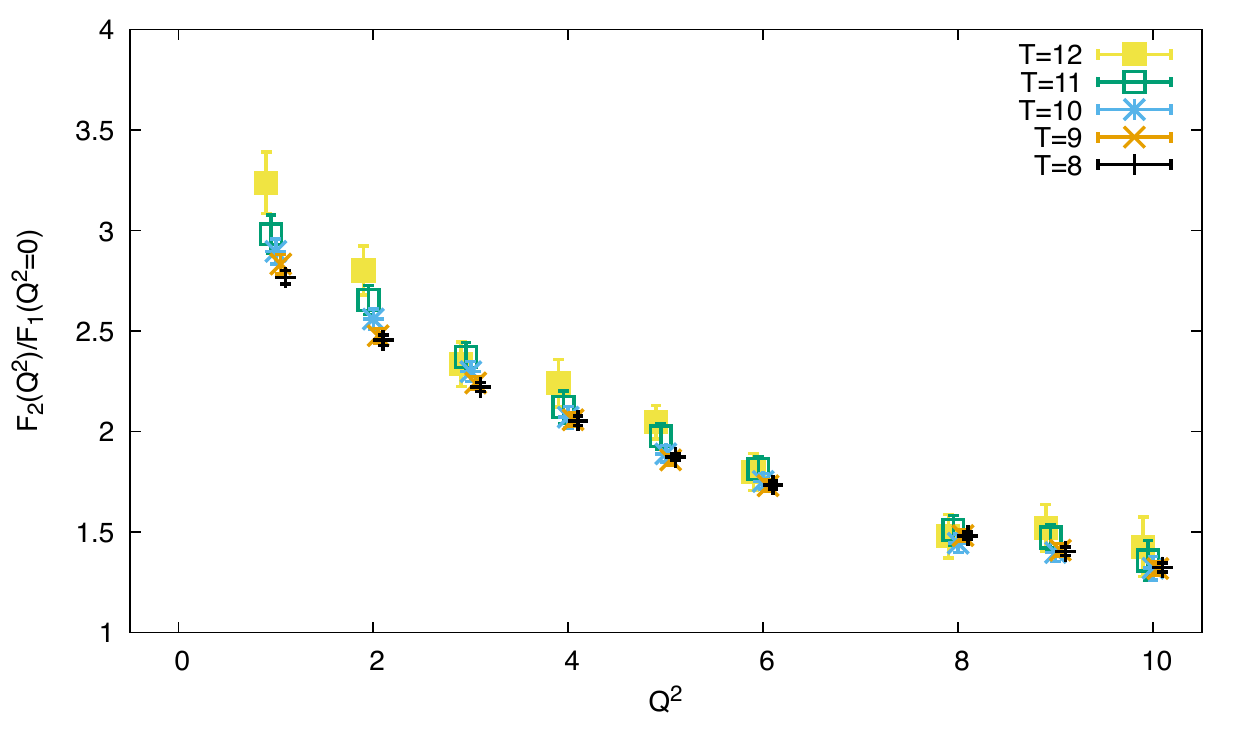}
\caption{
\label{fig:F2scaled}
Isovector induced tensor form factor, \(F_2(Q^2)\), scaled by the corresponding charge, \(F_1(0)\).
The shape does not depend on source-sink separation \(T\).
These extrapolate to \(\sim 3.2(2)\).  The experiment is 2.7928473446(8) + 1.9130427(5) - 1 = 3.705874(5) \cite{Workman:2022ynf}.
}
\end{center}
\end{figure}
\begin{figure}[b]
\begin{center}
\includegraphics[width=.7\textwidth,clip]{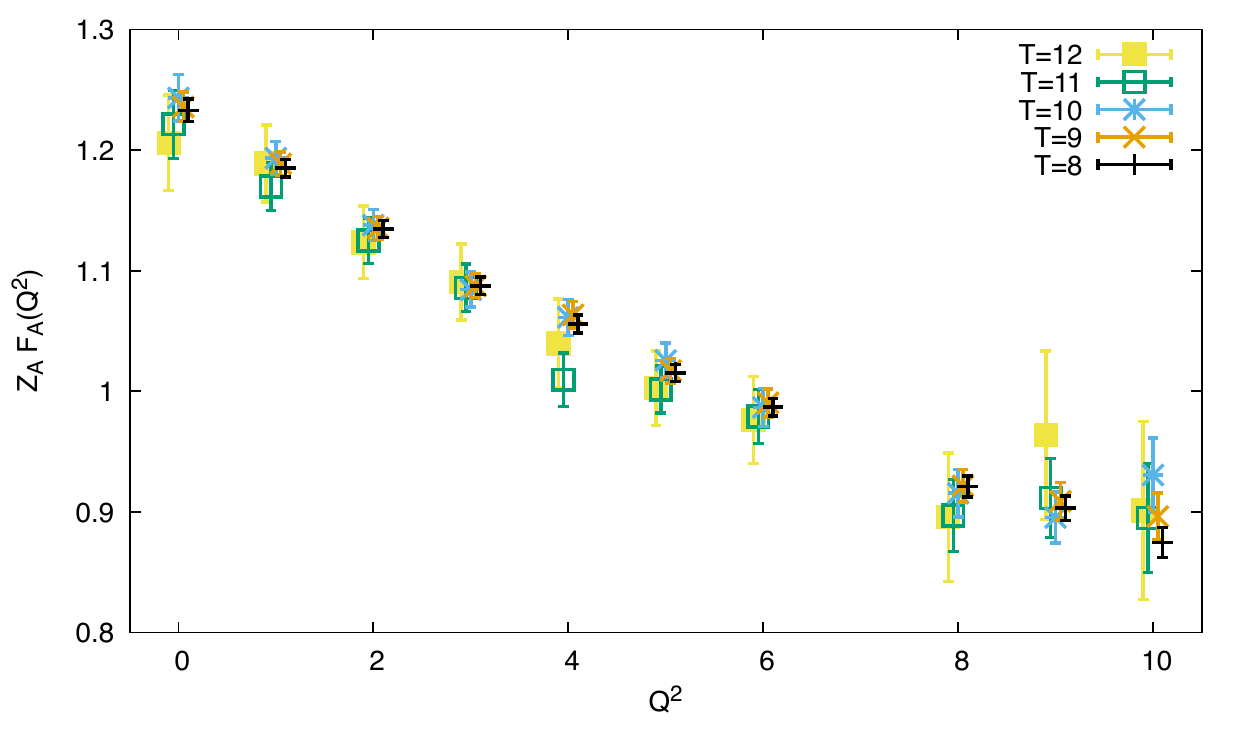}
\caption{
\label{fig:FAscaled}
Isovector axialvector form factor, \(F_A\), of the axialvector current, renormalized with \(Z_A=0.71191(5)\) obtained in the meson sector \cite{RBC:2014ntl}.
These extrapolate to \(\langle r_A^2\rangle \sim 0.20(2) \mbox{\rm fm}^2\).
}
\end{center}
\end{figure}

Again the shape is not affected by excited states either in the sense that the values calculated with shorter separations are well contained within the error bars of the values calculated with longer separations.

This form factor also is well fitted by the same wide range of multipole forms with  \(1 \le p \le 7\).
The resulting extrapolation estimates for the isovector anomalous magnetic moment, \(F_2(0)\sim 3.2\), do not depend on the multipolarity in this range.
Nor do they differ much from linear extrapolation using the two smallest available momenta transfer, \(Q^2=1\) and 2.
The multipole form with \(p\le 1/2\) or \(8 \le p\) does not work.

Axialvector form factor, \(F_A\), of the axialvector current is presented in Fig.\ \ref{fig:FAscaled}:
This is an important observable for the ongoing neutrino experiments but is poorly known only from bubble-chamber experiments in the 1970s.

The induced pseudoscalar form factor, \(F_P\), of the axialvector current is presented in Fig.\ \ref{fig:FPscaled}:
\begin{figure}[tb]
\begin{center}
\includegraphics[width=.7\textwidth,clip]{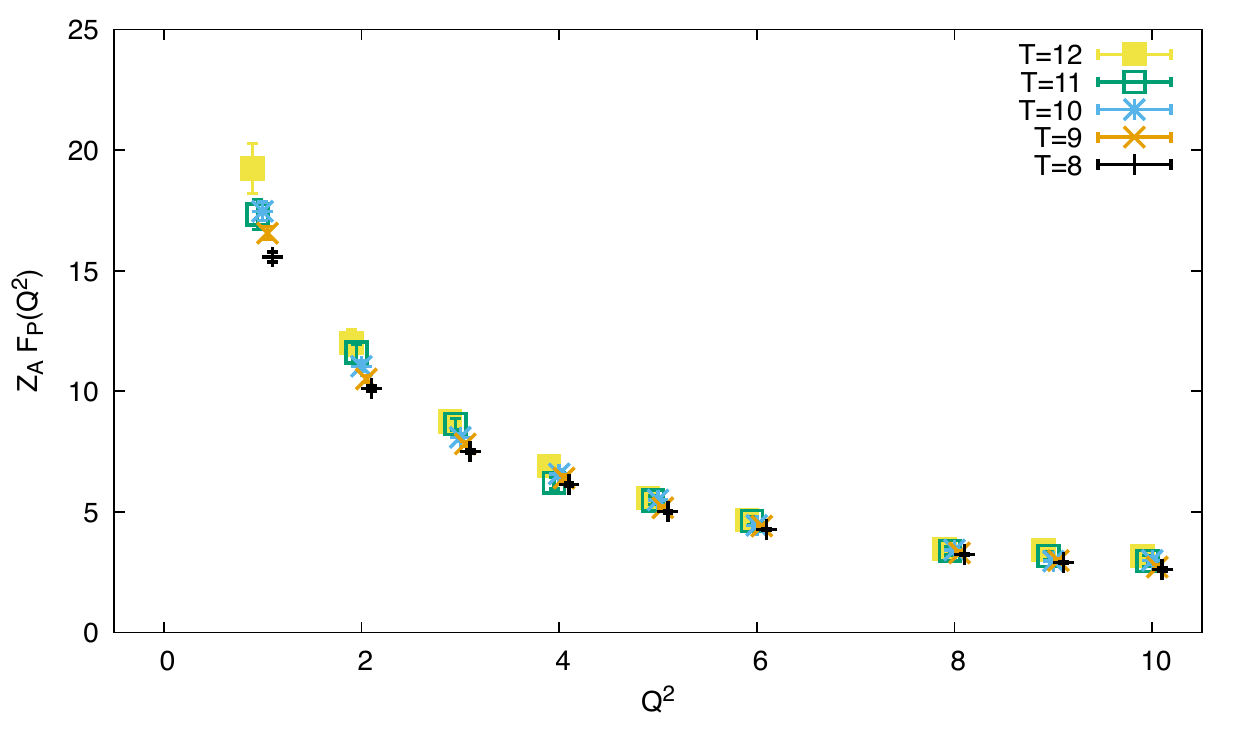}
\caption{
\label{fig:FPscaled}
Isovector pseudoscalar form factor, \(F_P\), of the axialvector current, renormalized with \(Z_A=0.71191(5)\) \cite{RBC:2014ntl}.
These extrapolate to \(F_P(0) \sim 25(2)\): likely result in much smaller \(g_P\) for muon capture.
}
\end{center}
\end{figure}
This is another crucial observable for processes such as muon capture.

Both these form factors are easily fitted by the same wide range of multipole forms with polarity \(1 \le p \le 7\).
The resulting mean-squared radius estimates of \(\langle r_A^2\rangle  \sim 0.18\ \mbox{\rm fm}^2\) or \(F_P(0)\sim 25\) do not depend much on the polarity \(p\).
Nor do they differ from the linear extrapolations using only the two smallest available \(Q^2\) values of either 0 and 1 or 1 and 2.
The multipole form with  \(p\le 1/2\) or \(8 \le p\) does not work.

In Table \ref{tab:comparison},
\begin{table}[t]
\begin{center}
\begin{tabular}{llllllll}\hline\hline
 & & \(T=8\) & 9 & 10 & 11 & 12 & experiment \\\hline\hline
\(\langle r_1^2\rangle\) \([\mbox{fm}^2]\) & linear&0.134(14) & 0.14(2) & 0.13(3) &  0.16(5) & 0.13(8) & 0.868(3)\\\
 & dipole & 0.135(6) & 0.143(8) & 0.142(13) & 0.14(2) & 0.13(3) & \\\hline
\(F_2(0)\) & linear & 3.159(4) & 3.250(6) & 3.242(8) & 3.252(13) & 3.61(2) & 3.705874(5) \\
 & dipole & 3.10(5) & 3.15(6) & 3.22(8) & 3.24(11) & 3.5(2) & \\\hline
\(\langle r_A^2\rangle\) \([\mbox{fm}^2]\) & linear & 0.177(2) & 0.174(2) & 0.182(4) & 0.192(5) & 0.066(8) & --\\
 & dipole & 0.177(7) & 0.174(10) & 0.176(14) & 0.18(2) & 0.15(3) & \\\hline
\(F_P(0)\) & linear & 21.01(3) & 22.61(5) & 23.90(7) & 23.04(11) & 26.5(2) & -- \\
 & dipole & 23(2) & 25(2) & 26(2) & 26(2) & 30(2)& \\
\hline\hline
\end{tabular}\\
\caption{
\label{tab:comparison}
The isovector form factor shape parameters obtained by dipole fits agree with those from linear extrapolations using only the smallest two \(Q^2\) values.
The vector-current parameters, however, disagree with well-established experiments \cite{Workman:2022ynf}.
The errors are single-elimination jack-knife statistical. 
}
\end{center}
\end{table}the isovector form factor shape parameters from the dipole fits are compared with those from the linear extrapolations using only the two smallest available \(Q^2=0\):
The form factor shape parameters from the dipole fits agree well with the corresponding linear extrapolations.
The corresponding extrapolations using other multipolarities of \(1 \le p \le 7\) do not differ.  
Consequently, the shape parameter estimates from other fit ansatzes, such as bounded \(z\) expansion, should not differ either, though we are yet to complete such analyses.

However, the vector-current form factor shape parameters we obtain here do not agree with the experiments.
We should investigate smaller momentum transfers than in the present study.
Such an investigation can be conducted with twisted boundary conditions in spatial directions for valence quark propagators.





The author thanks the members of LHP, RBC, and UKQCD collaborations, particularly Sergey Syritsyn.
The ``48I'' ensemble was generated using the IBM Blue Gene/Q (BG/Q) ``Mira'' machines at the Argonne Leadership Class Facility (ALCF) provided under the Incite Program of the US DOE, on the ``DiRAC'' BG/Q system funded by the UK STFC in the Advanced Computing Facility at the University of Edinburgh, and on the BG/Q  machines at the Brookhaven National Laboratory.
The nucleon calculations were done using ALCF Mira.
The author was partially supported by the Japan Society for the Promotion of Sciences, Kakenhi grant 15K05064.

\bibliographystyle{apsrev4-2}
\bibliography{nucleon}

\end{document}